\documentclass[12pt,preprint]{aastex}
\usepackage{CJK}
\shortauthors{Sadjadi et al.} 
\shorttitle{Alphatic side groups}

\usepackage{graphicx}
\usepackage{epsfig}

\begin{document}

%\title{Theoretical Insight to the MAON hypothesis: Addition of Aliphatic groups to the PAH fragment}

\title{A Theoretical Study on the Vibrational Spectra of PAH Molecules with Aliphatic Sidegroups
}
\begin{CJK*}{Bg5}{bsmi}
\CJKtilde

\author{SeyedAbdolreza Sadjadi, Yong Zhang (±iªa), Sun Kwok (³¢·s)}
\affil{Space Astronomy Laboratory, Faculty of Science, The University of Hong Kong, Pokfulam Road, Hong Kong, China}
\email{sunkwok@hku.hk} 

\begin{abstract}

The role of aliphatic side groups on the formation of astronomical unidentified infrared emission (UIE) features is investigated by applying the density functional theory (DFT) to a series of molecules with mixed aliphatic-aromatic structures.  The effects of introducing various aliphatic groups to a fixed polycyclic aromatic hydrocarbon (PAH) core (ovalene) are studied.    Simulated spectra for each molecule  are produced by applying a Drude profile at $T$=500 K while the molecule is kept at its electronic ground state.  The vibrational normal modes are classified using a semi-quantitative method.   This allows us to separate the aromatic and aliphatic vibrations and therefore  provide clues to what types of vibrations are responsible for the emissions bands at different wavelengths.  We find that many of the UIE bands are not pure aromatic vibrational bands but may represent coupled vibrational modes.  
The effects of aliphatic groups on the formation of the 8 $\mu$m plateau are quantitatively determined. The vibrational motions of methyl ($-$CH$_3$) and methylene ($-$CH$_2-$) groups can cause the merging of the vibrational bands of the parent PAH and the forming of broad features. %The analysis also provides a better understanding of the types of vibrations under the UIE bands especially at long wavelengths.
These results suggest that aliphatic structures can play an important role in the UIE phenomenon. 

\end{abstract}

\footnote {\today}

\maketitle
\end{CJK*}

\section{Introduction}

While it is widely known that interstellar dust contains an aromatic component,  the presence of an aliphatic component is  less well appreciated.   
The 3.4 $\mu$m feature was first detected in absorption against the infrared background of the galactic center  \citep{soifer1976}, but no identification was made at that time.  This feature was rediscovered by \citet{wick1980} a few years later in a search for organic dust in the diffuse interstellar medium.  Higher spectral resolution observations of the 3.4 $\mu$m feature in the planetary nebula IRAS 21282+5050 show that it is composed of several subfeatures at 3.4, 3.46, 3.51, and 3.56 $\mu$m  \citep{jou90}, which can be identified as C$-$H stretching modes of methyl ($-$CH$_3$) and methylene ($-$CH$_2$) groups in aliphatic hydrocarbon materials \citep{duley1983}. Since then, the feature has been seen in emission in proto-planetary nebulae \citep{geballe1992, hrivnak2007} and in absorption in galaxies \citep{wright1996, dartois2004, dartois2007}.

In addition to the 3.4 $\mu$m aliphatic C$-$H stretching mode, features at 6.9 and 7.3 $\mu$m corresponding to the bending modes of aliphatic methyl and methylene groups have been seen as early as 1979 \citep{willner1979}.  These features have also been detected in absorption \citep{chiar2002} as well as in emission \citep{kwok1999}.  

{\it Akari} observations of M82 have shown that the strength of the 3.4 $\mu$m feature increases from the disk to the halo of the galaxy, suggesting that aliphatic organics are distributed throughout the volume of galaxies \citep{yamagishi2012}. Since abundance can easily be derived from the strengths of absorption features, it is estimated that the amount of carbon tied up in aliphatic compounds represent $15-30\%$ of the total interstellar carbon \citep{dartois2011}. 

Another manifestation of the aliphatic component is the strong, broad emission plateau features seen in planetary nebulae and proto-planetary nebulae.  The plateau features centered around 8 and 12 $\mu$m were first seen in the {\it IRAS LRS} \citep{kwok1989} and the {\it Kuiper Airborne Observatory} \citep{buss1990} spectra of proto-planetary nebulae.  These two plateau features are much better defined by {\it ISO} observations and have been identified as superpositions of a variety of aliphatic side groups attached to aromatic rings \citep{kwok01}.  These broad features cannot be explained by the traditional pure aromatic models such as the polycyclic aromatic hydrocarbon (PAH) molecules.  The presence of these features, as well as the 3.4, 6.9, and 7.3 $\mu$m features, suggests that the chemical structures of organic dust in the interstellar medium is much more complex than previously believed.  

By introducing H into graphite ($sp^2$) and diamond ($sp^3$), a variety of amorphous C$-$H alloys can be created \citep{robertson2002}.  
Different geometric structures with long- and short-range can be created by varying the aromatic to aliphatic ratios. The spectral characteristics of these structures and their possible role as carrier of the unidentified infrared emission (UIE) bands have recently been extensively studied \citep{jones2012a, jones2012b, jones2012c}.  Laboratory studies of such amorphous carbonaceous solids \citep{dischler1983a, dischler1983b} suggest that they show many similar spectral characteristics as astronomical UIE bands. %\citep{kwok2014}.

The vibrational properties of PAH molecules have been extensively studied \citep[e.g.,][]{baus1997, ricca2013}. Infrared spectra  of PAHs  with different geometries, sizes, and charge states have been obtained with quantum chemical calculations \citep{malloci, baus2010, boersma2014}. 
%Although the vibrational properties of PAH molecules have been extensively studied \citep{baus1997, ricca2013}, very little is known about the vibrational properties of mixed aromatic/aliphatic carbonaceous compounds.
Recently, there has been an increasing interest in the effects of aliphatic structures.  The UIE bands have been suggested to be better explained by mixed aromatic/aliphatic organic nanoparticles \citep[MAON,][]{kwok2011}. The contributions of methyl and methylene groups have been computed by \citet{yang2013}, and the vibrational spectra of a variety of aliphatic structures have been calculated by \citet{papoular2014}.   %In this paper, we explore the spectral effects of aliphatic side group attached on aromatic clusters.

%MAONs molecular structure has been proposed as a very large 3D amorphous structure composed of benzene rings connected by aliphatic groups[3]. The molecule could possibly have thousands of carbon atoms as well as heteroatoms such as oxygen and sulfur[3]. Such  amorphous structures such as diamond-like amorphous carbon[4] or its hydrogenated form i.e. a-C:H[5] are known in carbon chemistry and have been studied by infrared spectroscopy till 1983[5, 6]. 
%Both MAON and its precursor structure  are classical organic chemistry structures in which all the carbon atoms possess $sp$, $sp^2$ or $sp^3$ hybridizations without any anomalous C$-$H bonds. In spite of the simplicity in the bonding patterns, conducting experimental studies on a large number of conformations is not feasible.  
Through advances in computational chemistry methods and the increasing power of  computing hardware, it is now possible  to perform theoretical investigation of the vibrational properties of large organic molecules \citep{papoular2013, ricca2013}.  With such theoretical studies, we can identify  the explicit roles of aliphatic groups in the vibrational spectrum of the molecules. In this work, we aim to study the changes in the infrared emission spectra of a PAH molecule upon addition of the aliphatic groups and to specifically identify the types of vibrational motions responsible for such changes.

\section{Computational methods}
   
Our calculations are based on  the Density Functional Theory \citep[DFT,][]{hohenberg1964, kohn1965, ziegler1991} where the molecular non-relativistic Schr\"odinger equation is solved under the Born-Oppenheimer approximation \citep{born1927}. The BH\&HLYP hybrid functional \citep{becke1993} in combination with polarization consistent basis set PC1 \citep{jensen2001, jensen2002} are applied to obtain  the equilibrium geometries and  the fundamental vibrational frequencies of hydrocarbon molecules. The chosen PC1 basis set includes the necessary polarized $d$ and $p$ orbitals on C and H centers individually. The accuracy of the vibrational frequencies under the BH\&HLYP/PC1 model is estimated to have an rms error of 2 cm$^{-1}$  \citep{lau12}. %In addition to this desirable accuracy the moderate size of the basis set enabled the calculations to be performed within reasonable computational time.  
%As a starting example, we study the different isomers of C$_{55}$H$_{52}$ molecule used by \citet{kwok01} as possible candidates responsible for plateau features in the observed UIE spectra. 
All geometries have been optimized under the default criteria, i.e. forces $< 4.5 \times 10^{-4}$ hartree/bohr and rms error of the forces less than $3.0 \times 10^{-4}$. Fine integration grid in combination with the default convergence criterion of $1 \times 10^{-8}$ is applied for self-consistent field calculation of DFT density matrix. The optimized geometries are all characterized as local minima, established by the positive values of all frequencies and their associated eigenvalues of the second derivative matrix. The vibrational frequencies are calculated via the harmonic normal mode approximation \citep{och99}. The scaling factors of 0.9311 and 0.9352 for vibrational frequencies $>1000$ cm$^{-1}$ and $<1000$ cm$^{-1}$, respectively, are then applied \citep{lau12}. The final infrared spectra for each isomer are constructed by plotting the scaled frequency values versus the calculated vibrational mode intensities.

The simple quantitative decomposition normal mode analysis is adopted in combination with
customary visualizing of the normal modes. The Cartesian displacements $x, y, z$ of all the atoms calculated in the vibrational analysis \citep{och99} are converted to a quantitative measure $R$, which is defined as

\begin{equation}
R=\sqrt{x^2+y^2+z^2}
\label{normal}
\end{equation}
where $x, y, z, R$ are all expressed in units of \AA.

The analysis is done with a small code called Vibanalysis written by SAS. %\citep{sad13} 
This code has the ability to read the atoms displacement in cartesian coordinates and to perform the described analysis. Its output is the list of the atoms (labeled as aliphatic or aromatic) with the major contribution in the corresponding normal mode, i.e. the atoms with $R\geq 0.1$ \AA.

All ab initio calculations have been performed by Gaussian 09, Revision C.01 package \citep{fri09}. The HKU grid-point supercomputer facility equipped with 8 and 12 CPU cores and 30$-$40 GB of RAM on each node has been utilized for this purpose. Visualizing and manipulating the results of vibrational normal mode analysis were performed by utilizing the Chemcraft suit program \footnote{http://www.chemcraftprog.com}. 

%\subsection{{Simulation of IR emission spectra}}

%In this section, we model the IR spectra of the molecules by applying the broadening model to search for plateau features. This has been explored previously by others via complex electronic excitations coupled with vibrational motions of the molecules (neutral or charged). Here we assume that all our neutral molecules will remain in their electronic ground state and the broadening is controlled by the thermal excitation of the vibrational modes.     

In order to facilitate comparison with the observed astronomical infrared emission spectra, the calculated  vibrational-mode intensities have been converted to a simulated observed spectra by applying a thermal excitation and a Drude profile.   
Assuming a thermal population, the intensities of emission lines can be
calculated from the vibrational-mode intensities calculated by DFT. We assume that the intensity of the emission line at certain wavelength is equal to its absorption intensity calculated by DFT. Ignoring the contributions from hot bands and overtones, the line intensity of the $i$th vibrational mode ($I_i$) is proportional to  $exp(-h\nu_i/kT)$ where $\nu_i$  is the frequency of the normal mode, and $h$ and $k$ are the Planck and Boltzmann constants, respectively. A Drude profile 
   
\begin{equation}
F_i(\lambda)= \frac{2I_i\lambda_i}{\pi c\gamma}\times\frac{\gamma^2}{(\lambda/\lambda_i-\lambda_i/\lambda)^2+\gamma^2}
\label{drude}
\end{equation}
is applied to broaden all the IR features, where $\lambda_i$  is the central wavelength,  $\gamma$ is  the  fractional full-width at half maximum (FWHM) of the simulated peaks and $c$ is the speed of light. The value of $\gamma$  chosen to fit the experimental or observational spectra. The modeled spectra are then obtained by sum over all $F_i(\lambda)$  values:

\begin{equation}
F(\lambda)=\Sigma F_i(\lambda)
\label{intensity}
\end{equation}
where $ F(\lambda) $ is the final intensity of the emission band at each point on the Drude profile.   In this model no electronic excitations have been considered therefore all the molecules in our present study are kept in their singlet electronic ground state.  $T=500$ K is chosen for the simulation temperature of emission spectra as this is the optimum temperature suggested by \citet{cook1998} that keeps the long wavelength vibrational levels well populated.

\section{Theory versus experiment}
\label{fitting}

%In order to estimate the error of chosen DFT model, we broadened the calculated spectra using the Drude function with the FWHM equal to that of experimental spectra (Fig.1 ?). 
%Each simulated IR bands is a composition of multiple vibrational transition lines. We measured bands peak positions and the corresponding normalized intensities as listed in Table1.
%The experimental data in Table1 are the peak positions and strength of the dominant band at each region of spectra ( Sandford 2013).

\begin{figure*}
\epsfig{file=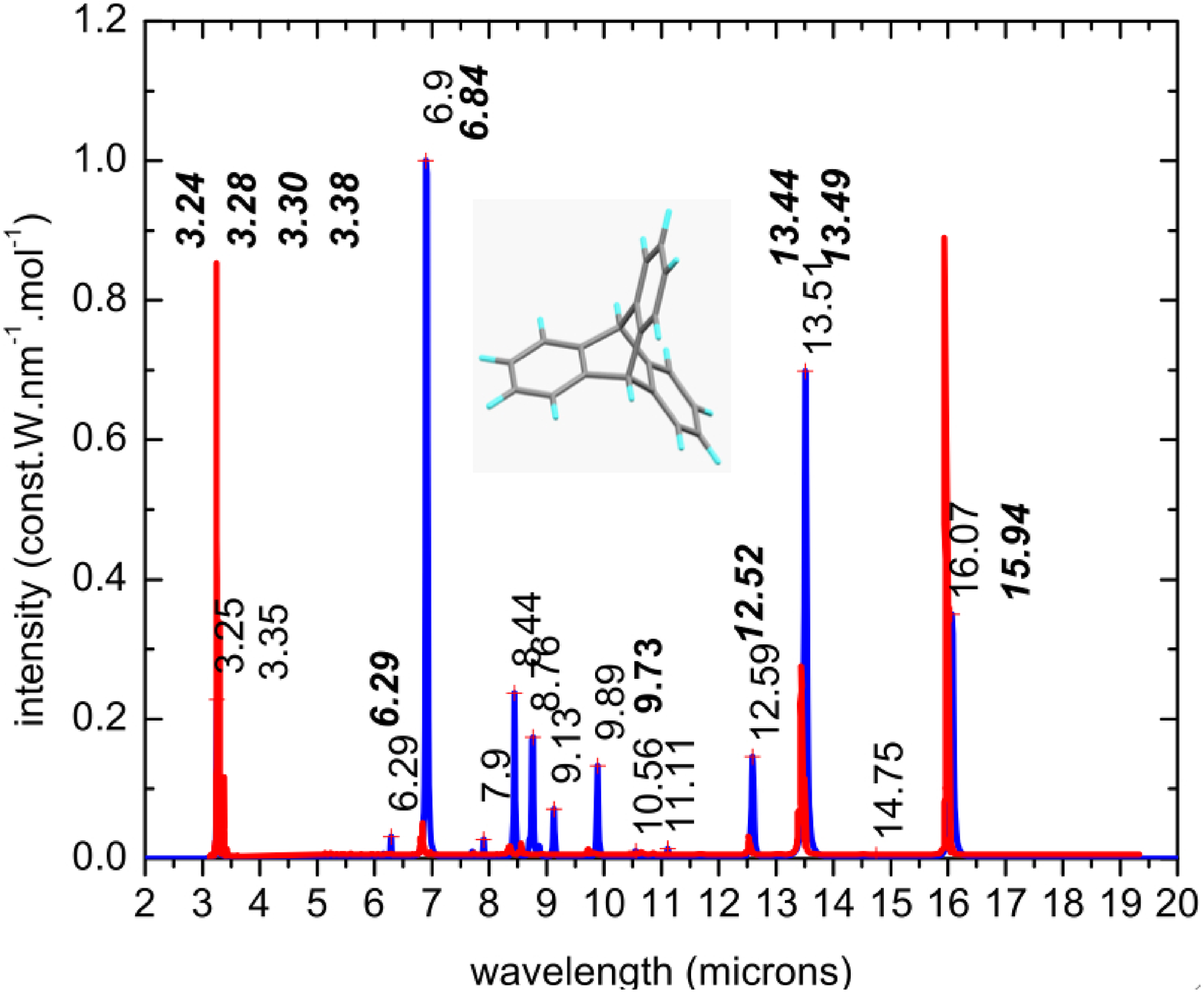, height=13cm}
%\plotone{figure1.eps}
\caption{The simulated IR spectra of C$_{20}$H$_{14}$ (shown in blue) at 500 K compared with the experimental data of \citet{sandford2013} (shown in red).  The broadening parameters were chosen to match to experimental broadening. The peak wavelengths of some of the features are also marked: theoretical values in normal font and experimental values in bold italics.  The optimized geometry of the molecule is shown in the insert. (A color version of this figure is available in the online journal)
\label{pah}}
%C100H118N4O15S4 
\end{figure*}

%\subsection {The Accuracy of the calculated IR lines}
In order to check the accuracy of our quantum chemical model in predicting the frequencies and intensities of the fundamental vibrational normal modes in hydrocarbon molecules, we compare our theoretical IR spectra with the recent experimental FT-IR data on number of hydrogenated PAH molecules reported by \citet{sandford2013}. These are the Group H molecules: C$_{22}$H$_{16}$(DHDB[ah]A), C$_{20}$H$_{14}$ (DHB[e]P) and C$_{20}$H$_{14}$(TRIP), the largest molecules in the set. %The calculated IR spectra and the optimized geometries are shown in Figure~\ref{pah}.
Qualitatively we found very good agreement between the calculated bands position and intensity ratios and  their corresponding experimental counterparts in the IR spectra.   A comparison between the calculated IR spectrum with the experimental spectrum of TRIP is shown in Figure \ref{pah}.  We have also evaluated the accuracy of our theoretical results quantitatively and the results are shown in Table 1.  The calculated spectra are broadened using the Drude function with the FWHM equal to that of experimental spectra. Each peak is actually a composition of multiple transitions. The peak positions and integrated strengths are measured and listed in Table 1.   
We find that the BH\&HLYP vibrational modes intensity ratios are in very good agreement with the reported experimental band strengths. 
%Accuracy of  $<$10 cm$^{-1}$ is achieved for bands with peak positions at 626.2 cm$^{-1}$ (16.0 $\mu$m),  742.4 cm$^{-1}$ (13.5 $\mu$m), 796.7 cm$^{-1}$(12.6 $\mu$m), 1193.2 cm$^{-1}$ (8.4 $\mu$m) and two calculated frequencies of the 3081.5 cm$^{-1}$ (3.2 $\mu$m) band.  For the peak positions of the other bands, we find an accuracy of 10 cm$^{-1}$ to 30 cm$^{-1}$, with the largest theoretical error found for the C$-$H stretching band at 2958.3 cm$^{-1}$ (3.4 $\mu$m) in Table 1.
Accuracy of  less than 0.2 $\mu$m is achieved for peak positions of the bands  with the the largest theoretical error found for 8.8 $\mu$m band in Table 1.
%\subsection{Reliability of the Model}

\begin{figure}
\epsscale{0.9}
\plotone{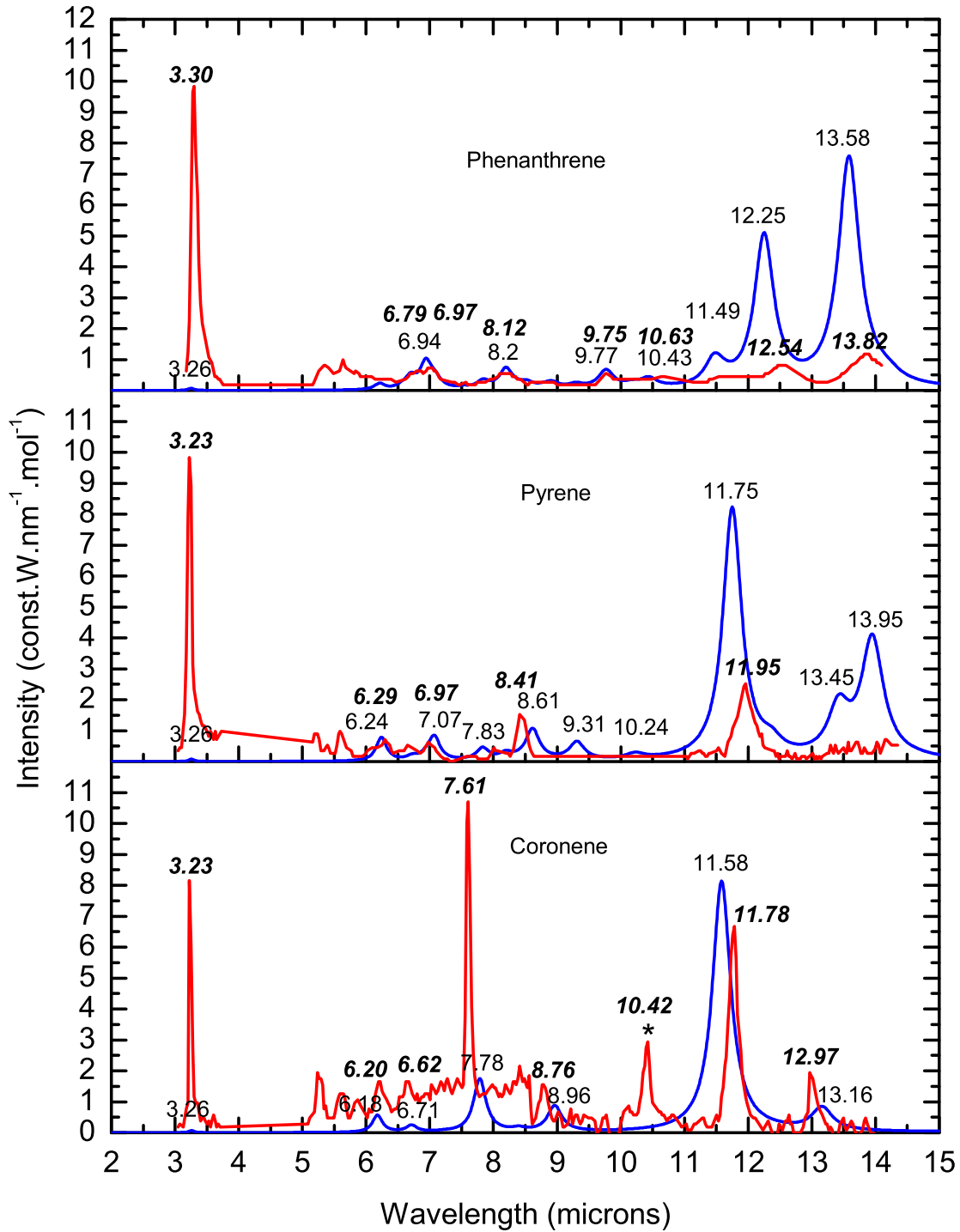}
\caption{Comparison of the experimental spectra (in red) of phenanthrene (top panel), pyrene (middle panel), and coronene (bottom panel) \citep{cook1996} with our theoretical DFT spectrum (in blue).  The numbers are peak wavelengths of the features with the experimental values in bold italics and the theoretical values in regular font. The * marks a contamination band in the spectrum of coronene \citep{cook1998b}. The digitized experimental spectrum intensity values have been scaled up by a factor of 10 for ease of comparison.  (A color version of this figure is available in the online journal)}
\label{pah_cook}
\end{figure}

The reliability of our method to simulate the infrared emission spectrum was verified against experimental spectra of PAH molecules \citep{cook1996}.
Comparisons are made with three selected PAH molecules, phenanthrene (C$_{14}$H$_{10}$), pyrene (C$_{16}$H$_{10}$) and coronene (C$_{24}$H$_{12}$)  (Figure \ref{pah_cook}). While there are differences in relative intensities, the positions of the theoretical bands are in good agreement with experimental values. The average absolute error of the position of simulated bands are 15.31 cm\textsuperscript{-1}, 20.68 cm\textsuperscript{-1} and 20.47 cm\textsuperscript{-1} (equivalent to 0.12 to 0.13 $\mu$m error) for phenathrene, pyerene and coronene, respectively.  The 10.42 $\mu$m  band in experimental spectra of coronene (Figure \ref{pah_cook}) is a contamination band as reported  by \citet{cook1998b}.

\section{Interpretation of emission spectra}

\begin{figure}
\epsscale{1.0}
\plotone{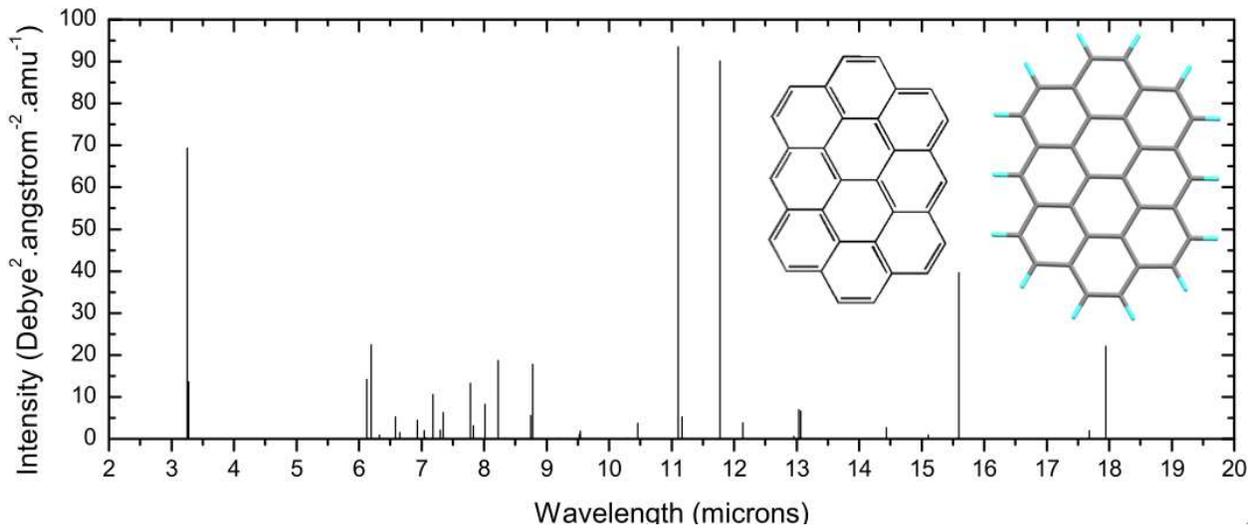}
\caption{Calculated 2$-$20 $\mu$m spectrum of C$_{32}$H$_{14}$.  The 2D structure and 3D capped sticks representation of the optimized geometry of C$_{32}$H$_{14}$ are also shown.  
\label{c32h14}}
\end{figure}

\begin{figure}
\plotone{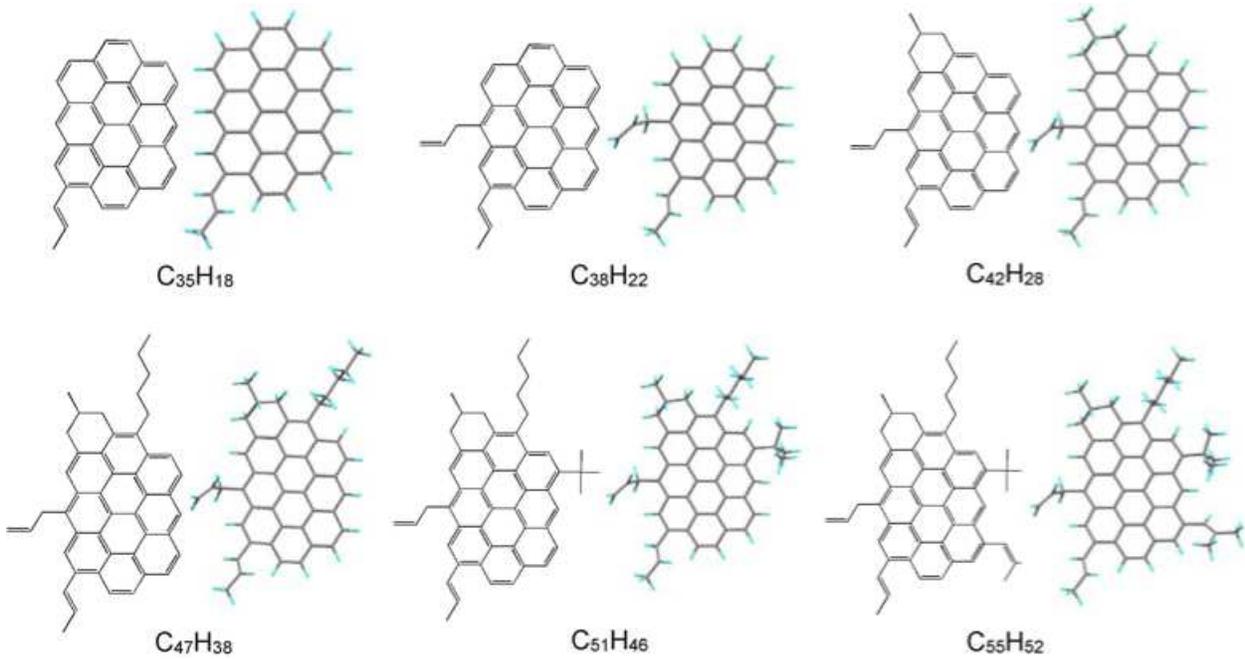}
\caption{The 2D and 3D capped stick representations of the optimized geometry of our sample molecules.  The molecules are (from left to right and from top to bottom) C$_{35}$H$_{18}$, C$_{38}$H$_{22}$, C$_{42}$H$_{28}$, C$_{47}$H$_{38}$, C$_{51}$H$_{46}$, and C$_{55}$H$_{52}$.  
\label{geometry}}
\end{figure}

The detection of astronomical aliphatic emission features has led to interest in the exploration of variations from the PAH model, which until recently has been the most widely accepted model for the astronomical UIE bands.  Recent efforts include the laboratory study of partially hydrogenated PAHs \citep{sandford2013} and theoretical studies of the vibrational modes of kerogen derivatives \citep{papoular2013}.  In this paper, we explore the intermediate case through the gradual addition of aliphatic side groups to an aromatic core.  We begin with a medium-size PAH molecule, ovalene (C$_{32}$H$_{14}$, Figure \ref{c32h14}), and gradually replace
its peripheral H atoms with the different $-$CH=CHCH$_3$, $-$CH$_2$CH=CH$_2$, $-$CH$_2$CH$_2$CH$_2-$, $-$CH$_2$(CH$_2$)$_3$CH$_3$, $-$C(CH$_3$)$_3$ and $-$CH=C(CH$_3$)$_2$ aliphatic functional groups.  Such structures are similar to those described in Figure 4 of \citet{kwok01}.
With different type of functional groups the symmetry of the molecule is reduced to $C_1$ and therefore the number of IR active modes is maximized.  The chemical formula, 2D structural formula, 3D ball and stick model for the local-minimum geometries of our sample molecules are shown in Figure \ref{geometry}.

%We showed that the major features in emission spectra can be reasonably simulated by Drude model.   Since the astronomical emission spectra is the primary source of molecular information, its important to show  how our theoretical tools help to extract such information.
In order to compare with astronomical spectra, we apply the Drude profiles to the molecules, creating broad emission bands from the vibrational lines. The normal mode vibrational line(s) that coincides with the peaks of  emission band are used to classify and label the band.  We find that in majority of the cases, the most intense IR lines under a particular emission band coincide well with the band peak position. For  bands that overlap with similar strength intensity IR lines, we rely on emission peak position to pick the corresponding vibrational mode.

The vibrational modes of the ovalene molecule in the 2$-$20 $\mu$m range are identified in Table \ref{ova}.  Each of the vibrational mode is assigned based on our analysis on displacement vectors  together with visual inspection of the animation of the vibrational modes. 
These results are generally consistent with usual interpretation of vibrations for aromatic/aliphatic compounds \citep{duley1983}. For example, the 3.3 $\mu$m band is assigned as aromatic C$-$H stretch.   However, some other bands are the result of coupled motions and their vibrational mode assignments are not as straight forward as the 3.3 $\mu$m feature.

\begin{figure}
\plotone{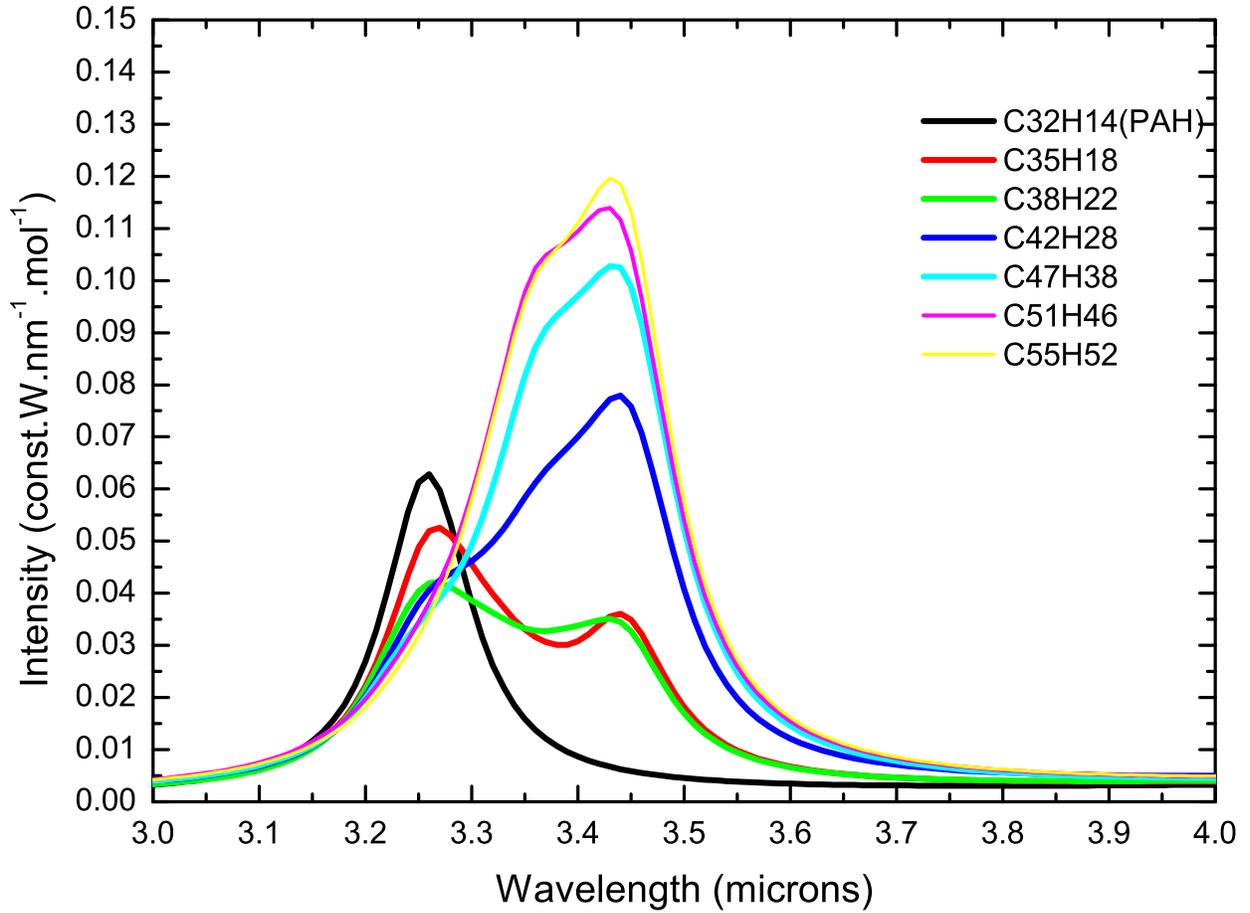}
\caption{Simulated 3.0$-$4.0 $\mu$m emission spectra of the 7 molecules with increasing degrees of aliphatic structures.  The first molecule C$_{32}$H$_{14}$ (ovalene) is a purely aromatic molecule..  (A color version of this figure is available in the online journal)
\label{expanded1}}
\end{figure}

\begin{figure}
\plotone{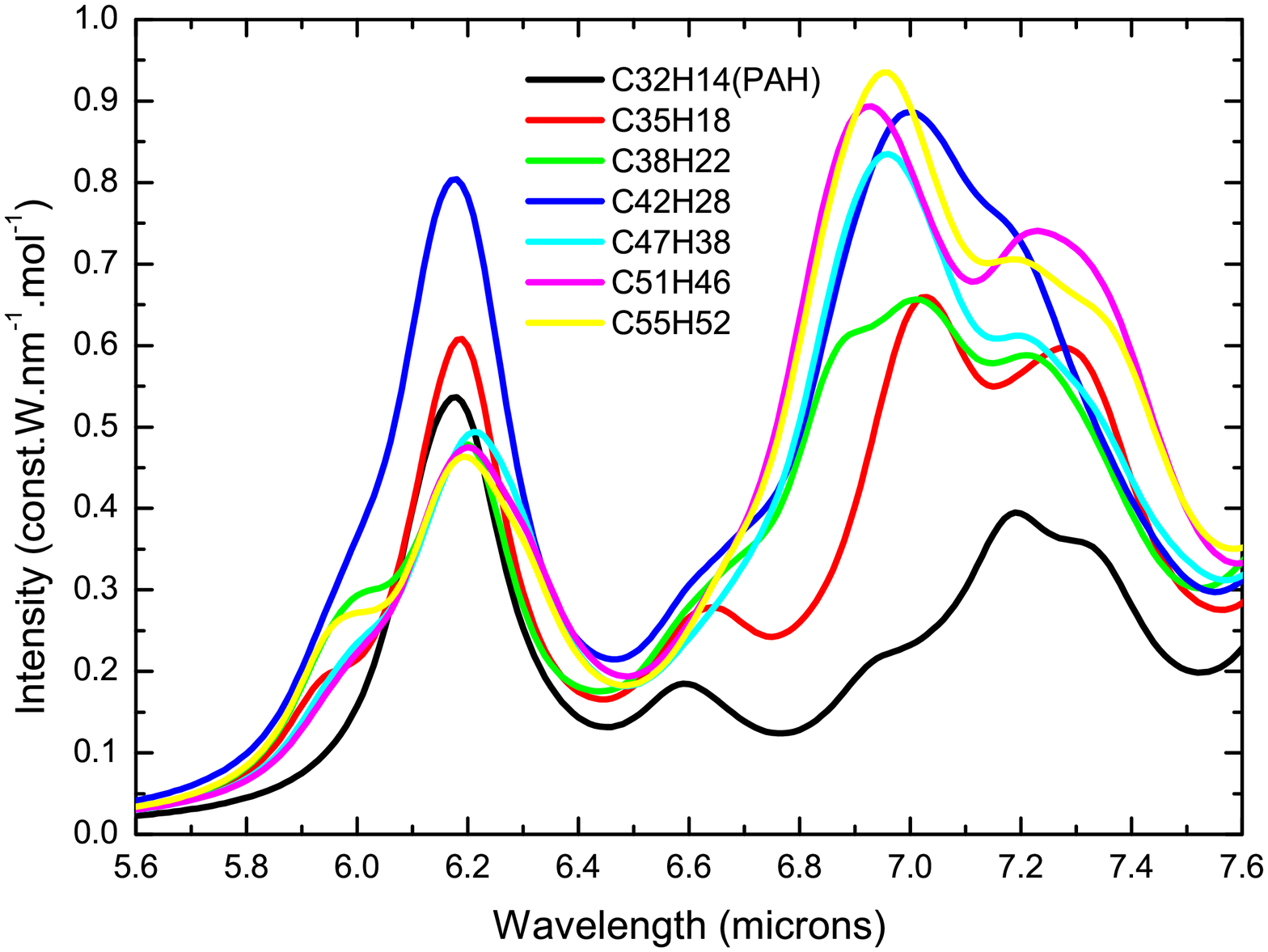}
\caption{Simulated 5.6$-$7.6  $\mu$m emission spectra of the 7 molecules with increasing degrees of aliphatic structures.  (A color version of this figure is available in the online journal)
\label{expanded2}}
\end{figure}

\begin{figure}
\plotone{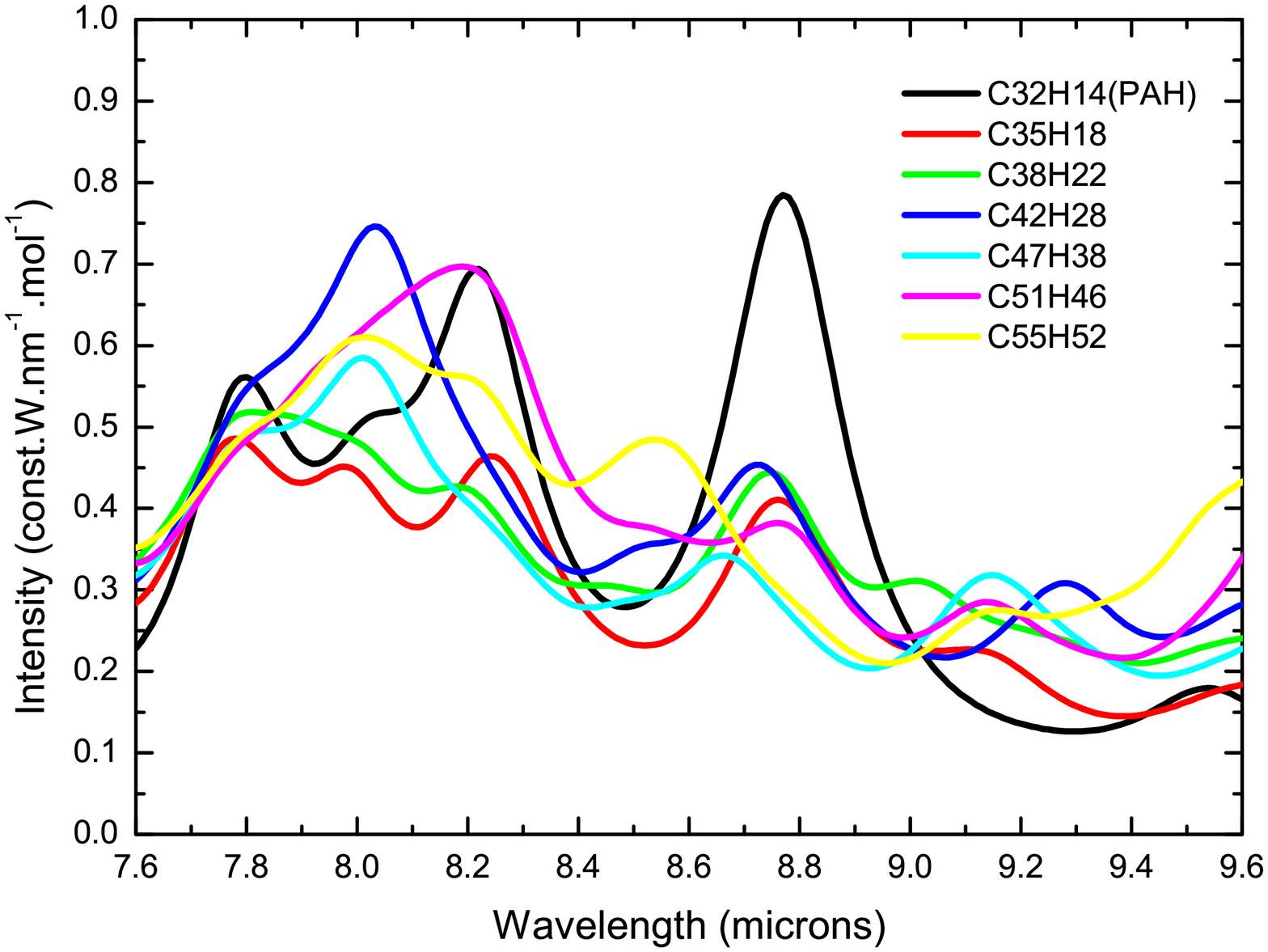}
\caption{Simulated 7.6$-$9.6  $\mu$m emission spectra of the 7 molecules with increasing degrees of aliphatic structures.    (A color version of this figure is available in the online journal)
\label{expanded3}}
\end{figure}

\begin{figure}
\plotone{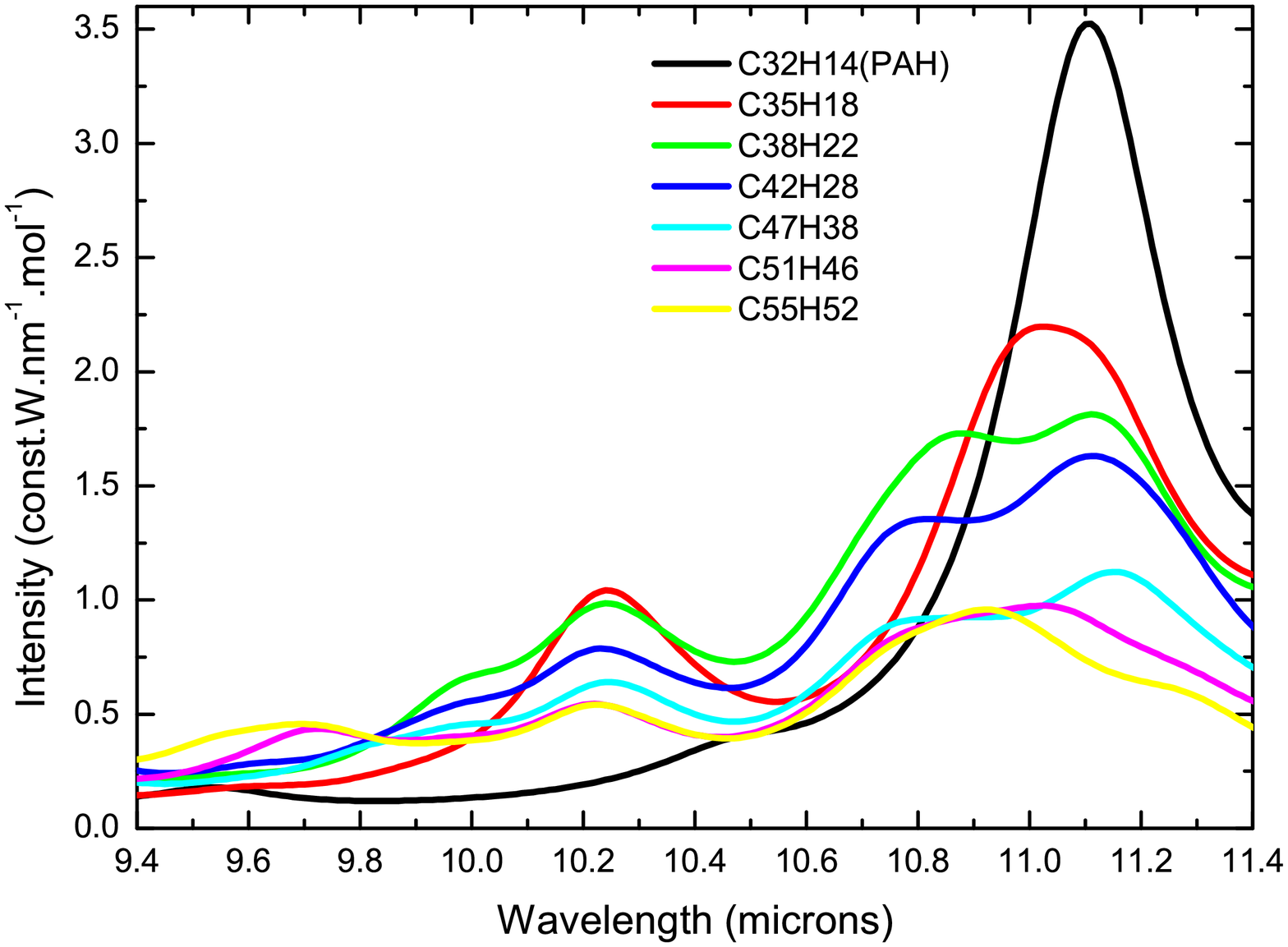}
\caption{Simulated 9.4$-$11.4  $\mu$m emission spectra of the 7 molecules with increasing degrees of aliphatic structures.   (A color version of this figure is available in the online journal)
\label{expanded4}}
\end{figure}

\begin{figure}
\plotone{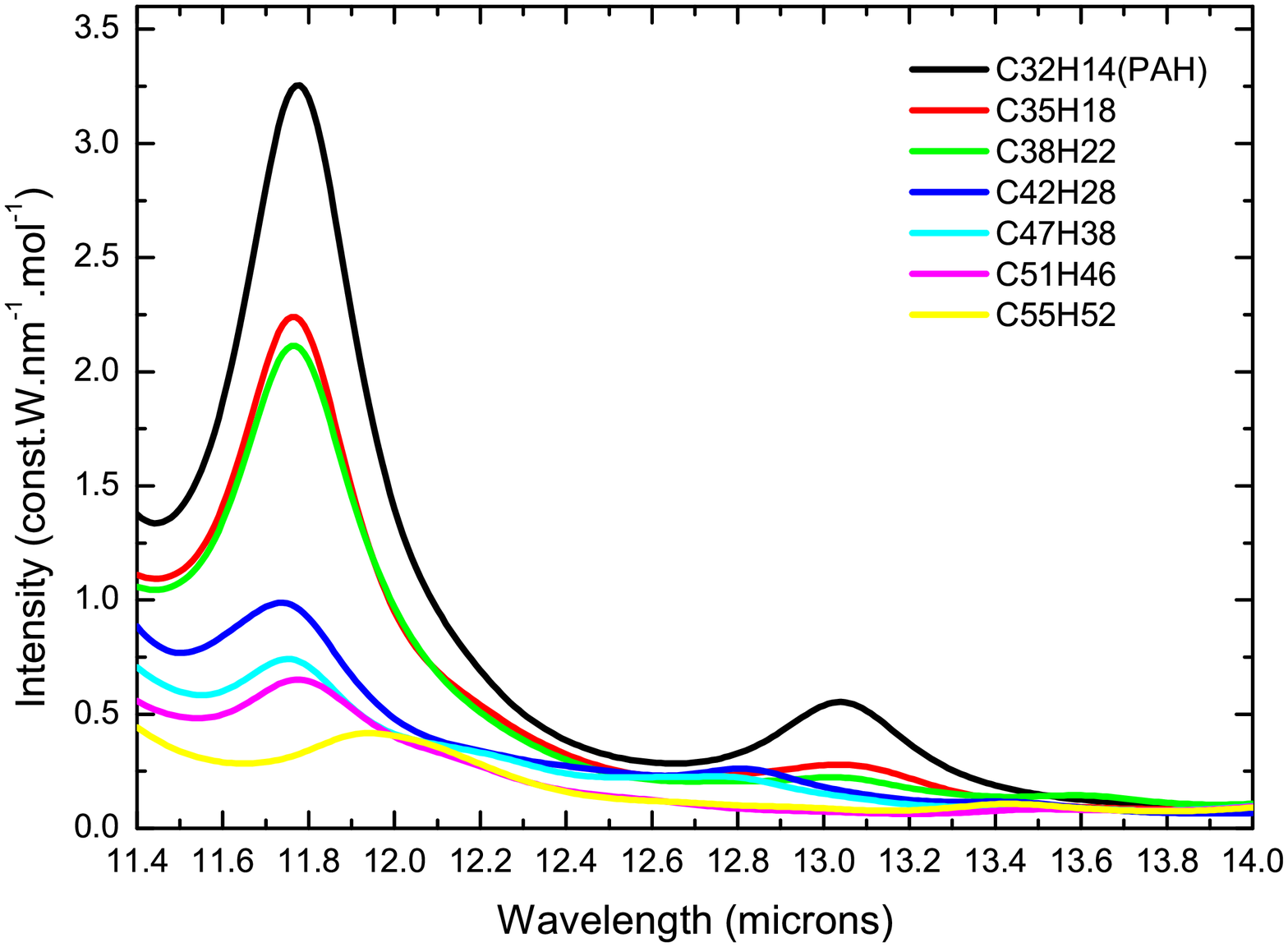}
\caption{Simulated 11.4-14.0   $\mu$m emission spectra of the 7 molecules with increasing degrees of aliphatic structures. (A color version of this figure is available in the online journal)
\label{expanded5}}
\end{figure}

\begin{figure}
\plotone{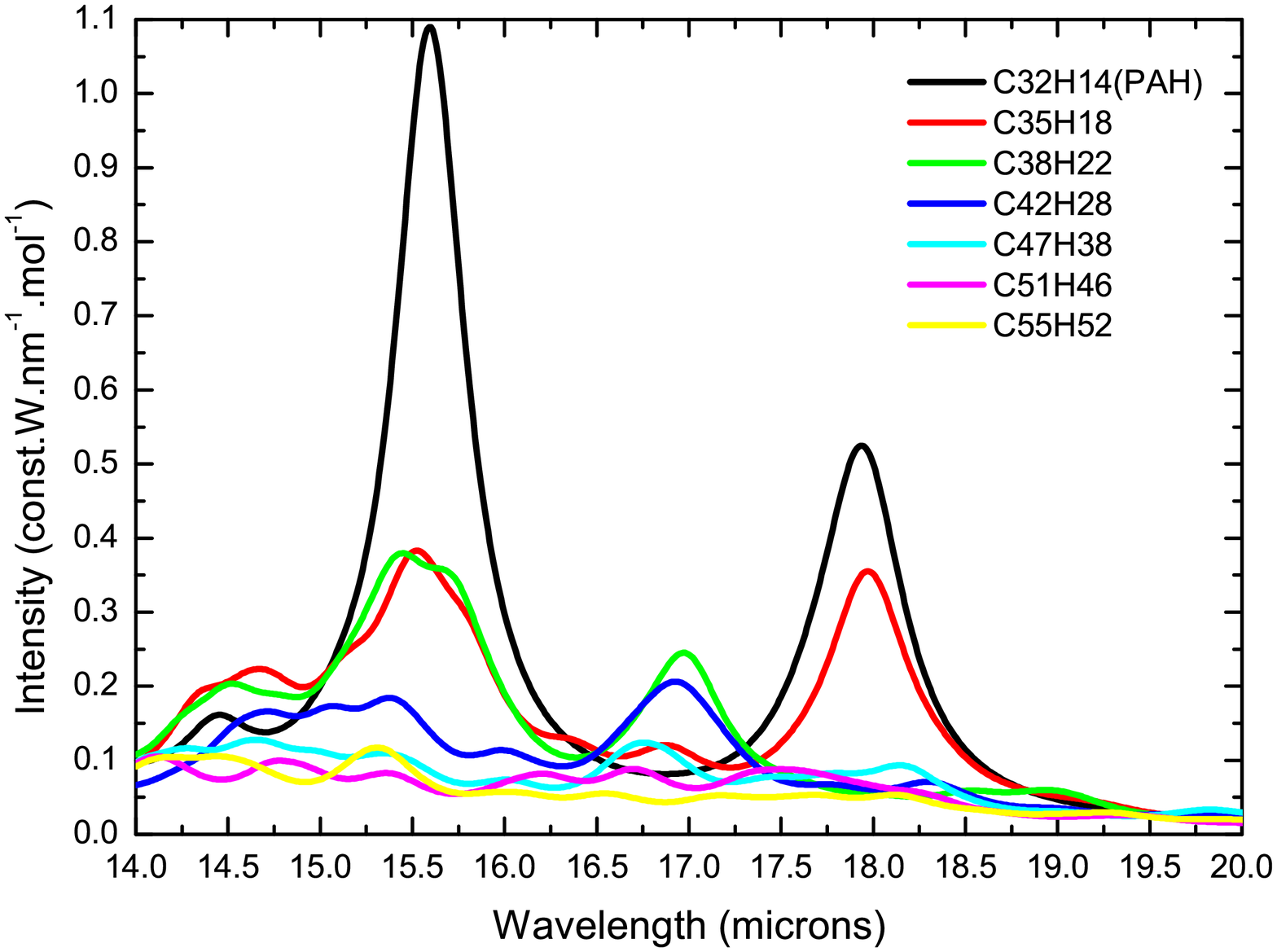}
\caption{Simulated 14.0$-$20.0 $\mu$m emission spectra of the 7 molecules with increasing degrees of aliphatic structures. (A color version of this figure is available in the online journal)
\label{expanded6}}
\end{figure}

%Since  our analysis provides more details on the vibrational motions under each emission band, we have extended the analysis to the other 6 molecules of mixed aliphatic/aromatic structure in our sample. 
Figures \ref{expanded1}$-$\ref{expanded6} show the spectral changes as aliphatic components are added to the ovalene molecule.    A glance at these figures show  that the PAH bands at 8.77 $\mu$m, 11.78 $\mu$m, 13.04 $\mu$m, 15.59 $\mu$m and 17.94 $\mu$m  weaken relatively upon the addition of aliphatic groups. Qualitatively, all of these bands show the contribution from in-plane or out-of-plane bending vibrations of aromatic C$-$H bonds (Table \ref{ova}).   The weakening of these bands is the result of replacing these hydrogens by aliphatic groups.

Below, we discuss in more details the spectra in separate wavelength regions.

\begin{itemize}
\item{3$-$4 $\mu$m region (Figure \ref{expanded1})

The ovalene molecule has a single emission band at 3.25 $\mu$m,  which is due to  aromatic C$-$H stretch. When aliphatic groups are added, a broader feature around 3.45 $\mu$m begins to appear.  
Vibrational analysis shows that this band is due to aliphatic ($sp^3$) C$-$H stretching modes.
The addition of more aliphatic groups causes the two bands to merge, as can be seen in the spectrum of C$_{55}$H$_{52}$.  These results are in agreement with functional group frequency assignment scheme widely used in organic chemistry \citep{fuhrer1972, gunzler}. }

\item{5.6$-$7.6 $\mu$m region (Figure \ref{expanded2})

The 6.2 $\mu$m band is a common feature among all the molecules under study. Normal mode analysis shows that this band is due to aromatic C=C stretching coupled with in-plane bending mode of aromatic C$-$H bonds.  An aliphatic feature is  observed as a weak peak at 6.01 $\mu$m  at the shoulder to the aromatic 6.2 $\mu$m band. 
This 6.01 $\mu$m feature can be identified as aliphatic C=C stretching coupled with aliphatic =C-H in-plane bending modes found in H$_2$C=CH$-$CH$_2-$ and H$_3$C$-$CH=CH$-$ functional groups. This band is blue shifted to 5.94 $\mu$m when the aliphatic double bond is conjugated to aromatic skeleton (C$_{38}$H$_{22}$). In this case the carbon atom of the ring bearing the aliphatic group also possess the considerable displacement and thus our normal mode analysis counts this motion as one of the mixed modes (C$_{38}$H$_{22}$). 
Ovalene (PAH) shows a weak band at 6.59 $\mu$m composed of coupled in plane C=C stretching  and C-H bending vibrations (Table \ref{ova}). The band shifts to 6.64 $\mu$m upon replacement of the first hydrogen atom by H$_3$C-CH=CH- group (C$_{35}$H$_{18}$). This band vanishes completely in all other molecules in our series. 

Aliphatic groups vibrational modes split the ovalene emission band at 7.19 $\mu$m into two bands. One centered at 7 $\mu$m and the other band at 7.25 $\mu$m. The vibrations at 7 $\mu$m are found to be deformation motions of $-$CH$_3$ (methyl) groups and scissoring motions of aliphatic $-$CH$_2$ (methylene) groups \citep{gunzler}. These aliphatic motions combine with aromatic vibrational modes at this region to form 7.25 $\mu$m band.  Addition of more aliphatic $-$CH$_3$ and $-$CH$_2$ groups results in merging these  two bands (7 $\mu$m and 7.25 $\mu$m) and the formation of  the broadened band at 6.95 $\mu$m (C$_{55}$H$_{52}$). This is also supported by the increase in the number of aliphatic and mixed modes as calculated via our normal mode decomposition analysis.}

\item{7.6$-$9.6 $\mu$m region (Figure \ref{expanded3})

In this spectral region, we see a marked growth in the number of pure aliphatic and mixed modes. As aliphatic groups are added to the ovalene molecule, three  distinct bands of ovalene at 7.80, 8.22, and 8.77 $\mu$m are merged together and appear as a one broad feature at around 8 $\mu$m in C$_{55}$H$_{52}$. Our vibrational analysis shows that the coupling of  $-$CH$_2$ groups twisting vibrational modes \citep{gunzler} with the aromatic vibrations at this region is the main reason for this change. Other types of aliphatic motions include H$-$C=C$-$H (para, in- plane bending), $-$CH$_3$ deformation, $-$C$-$H (aromatic, bending) modes. The strong 8.77 $\mu$m band in ovalene is blue shifted gradually upon combining with aliphatic vibrations and ultimately appears as weak band around 8.54 $\mu$m in the C$_{55}$H$_{52}$. 
 
The vibration under the weak 9.54 $\mu$m band of ovalene can not simply be assigned by usual terminology. We suggest using in-plane rocking mode of aromatic HC=CH unit for labeling this vibration. This band is replaced by the weak band in the range of 9.1$-$9.3 $\mu$m as a result of combination with $-$CH$_3$ and $-$CH$_2$ deformation, wagging and twisting modes. The band intensity (9.1$-$9.3 $\mu$m) increases when these aliphatic groups are added in the form of cyclohexane group (C$_{47}$H$_{36}$).  }

\item{9.4$-$11.4 $\mu$m region (Figure \ref{expanded4})

C$_{55}$H$_{52}$, the molecule with the largest number of aliphatic groups, shows a weak band at 9.7 $\mu$m . This band can be identified as coupling between various twisting/wagging modes of $-$CH$_2$ groups and deformation mode of $-$CH$_3$ groups and aromatic in-plane C$-$H bending modes. The vibration of aromatic C$-$H bonds are different from what we observed in the ovalene molecule at 9.54 $\mu$m, which weakens with the addition of aliphatic groups.

The out-of-plane bending modes of  para hydrogens in  H$_3$CHC=CH$-$ group  results in appearance of a band at 10.24 $\mu$m (see the spectrum of C$_{35}$H$_{18}$). This feature exists for all the molecules in our series. One of the other characteristic vibrations that appear under the 10.24 $\mu$m band is the stretching mode of the C$-$C single bond in the pentane side chain. 

The most notable feature in this spectral region is the strong band of ovalene (PAH) at 11.11 $\mu$m (Table \ref{ova}).  The combination of aromatic $-$C$-$H out-of-plane bending modes with deformation vibrations of $-$CH$_3$ groups shifts this strong band to 11.03 $\mu$m band, as can be seen in the spectrum of the C$_{35}$H$_{18}$ molecule. Addition of H$_2$C=CHCH$_2-$ aliphatic group causes the formation of another band at 10.85 $\mu$m. The aliphatic vibrations under this band are out-of-plane bending modes of olefin  C$-$H bonds (H$_2$C=). These vibrations are coupled with the out-of-plane bending modes of aromatic C$-$H bonds. By increasing the numbers of $-$CH$_3$ and $-$CH$_2$ groups, the deformation modes of $-$CH$_3$ groups and twisting or rocking modes of $-$CH$_2$ groups are added to the olefinic vibrations at this region of spectra. The coupling between all these aliphatic modes and the aromatic out-of-plane C$-$H bending modes results in the replacement of strong 11.11 $\mu$m feature in ovalene with the 10.92 $\mu$m broad band in C$_{55}$H$_{52}$. }

\item{11.4$-$14.0 $\mu$m region (Figure \ref{expanded5})

%The trends among the numbers of the aliphatic, aromatic and mixed modes in Table 2 exhibit the effects of aliphatic groups in our models are decreased beyond 12$\mu$m region. We anticipated that these effects are limited to activation of Raman modes in parent PAH rather than addition of new bands. However enlarging the size of aliphatic groups  may change this behavior.  

The strong 11.78 $\mu$m band (Table \ref{ova}) in ovalene is gradually weaken by the decrease in the number of aromatic C$-$H bonds in the other molecules. At the same time,  the band position is red shifted to 12 $\mu$m by combining the C$-$H out-of-plane bending mode with wagging and deformation modes of $-$CH$_2$ and $-$CH$_3$, especially the methyl and methylene group on the cyclohexane ring.  The coupling of various C$-$C$-$C$^{aromatic}$ bond angle deformation modes of aliphatic groups (for instance: H$_2$C=CH$-$CH$_2$-C$^{aromatic}$ with the PAH modes at 13.04 $\mu$m, forms a very weak band at around 13.34 $\mu$m in C$_{55}$H$_{52}$. This feature can  hardly be seen in Figure \ref{expanded5}.}

\item{14.0$-$20.0 $\mu$m region (Figure \ref{expanded6})

While the bands of ovalene (PAH) can be identified by the usual vibrational modes (Figure \ref{pah_movie}), the bands of the other molecules in our series in this part of the spectrum cannot be so easily labeled (Figure \ref{maon_movie}).  Our simple displacement vector analysis shows that the motions involve a large fraction of the molecules and are more complex in nature.  However, distinction between aromatic and aliphatic contributions can still be made.  
The ovalene (PAH) molecule has a weak band at 14.45 $\mu$m and two strong bands at 15.59 $\mu$m and 17.94 $\mu$m.  These bands are weaken by replacement of aromatic C$-$H bonds with aliphatic groups. The only noticeable fingerprint of the aliphatic group in this region is the concerted rocking motion of $-$CH$_2$ groups on the pentane side chain. This vibration forms a band at 15.31 $\mu$m in C$_{55}$H$_{52}$. Based on our analysis it is suggested that the strong 17.94 $\mu$m band (Table \ref{ova}) in ovalene is blue shifted to 16.9 $\mu$m region and then it gradually vanishes.  }

\end{itemize}

\section{Discussions}

The UIE phenomenon is a complex one.  In addition to the aromatic features at 3.3, 6.2, 7.7, 8.6 and 11.3 $\mu$m, there are aliphatic features at 3.4 and 6.9 $\mu$m.  There are also unidentified features at 15.8, 16.4, 17.4, 17.8, and 18.9 $\mu$m which are observed in proto-planetary nebulae \citep{kwok1999}, reflection nebulae \citep{sellgren2007}, and galaxies \citep{sturm2000}.  Broad emission plateau features at 8 and 12 $\mu$m, and maybe also around 17 $\mu$m are commonly observed.   In recent years, there has been an increasing interest in the plateau features as they are found to be associated with the presence of fullerene (C$_{60}$) \citep{zhang2013, otsuka2014}.  The unidentified 21 $\mu$m feature \citep{kwok1989} is also often accompanied by the 8 and 12 $\mu$m plateau features \citep{cerr2011, volk2011}.  Suggestions have been made that the carrier of the plateau features are precursors to fullerene \citep{ber2012, garcia2012}.   A good understanding of the origin of the plateau features is key to the identification of chemical structure of the carrier of the UIE bands and the history of chemical synthesis in the late stages of stellar evolution.

How well can a mixed aromatic/aliphatic model work in simulating the astronomical UIE bands in comparison to the PAH model? Traditionally, the UIE bands are explained by collective emissions from a mixture of PAH molecules. \citet{cook1998} used a family of nine neutral PAHs (including ovalene) to simulate the astronomical spectra. They found that the inclusion of cation PAHs can explain the UIE better than a collection of neutral PAHs, but their simulation still cannot reproduce the details of UIE.   They suggest that a larger number of PAHs and/or  larger PAHs are required.   With data on hundreds of PAH molecules now available from the NASA Ames PAH database \citep{boersma2014}, it is possible to use PAH molecules with different sizes and charged states to simulate the astronomical UIE spectra \citep{cami2011}.  However, questions have been raised on how meaningful such fitting exercises are due to the large number of parameters involved \citep{zhang2015}.
 
Although the range of mixed aromatic/aliphatic structures that we have calculated is limited, Figure \ref{cook} (C$_{55}$H$_{52}$, an ovalene molecule with aliphatic branches) shows that aliphatic side groups can also create a host of features in the 6$-$9 and 10$-$14 $\mu$m regions, and therefore offers the possibility of explaining the 8 and 12 $\mu$m plateau emission features. In the 15$-$20 $\mu$m region, the mixed aromatic/aliphatic structure can lead to complex vibrational motions, and could form the basis of explaining  the 17 $\mu$m plateau emission feature. These results suggest that the inclusion of aliphatic side groups will help  our interpretation of the UIE phenomenon.

\section{Conclusions}

%We introduced the set of theoretical tools to simulate the infrared emission spectra of single neutral organic molecules with moderate size and simple bonding patterns at singlet electronic ground state.  
%We have demonstrated that the resulting emission spectra can capture the essential features observed in the astronomical emission spectra  without taking the complex electronic excitation phenomena into account.
We have conducted a theoretical study on the changes on the infrared emission spectra of the PAH molecule ovalene through the replacement of aromatic hydrogen atoms by aliphatic groups.  The local-minimum geometry and the corresponding vibrational normal modes are calculated by applying the density functional theory.  The IR absorption lines positions are then corrected for anharmonicity, couplings and other complex electronic effects, by applying the most recently developed vibrational scale factor scheme. In order make comparison with astronomical spectra, these absorption lines are translated to emission bands assuming a Drude profile and thermal excitation. The resulted emission spectra show good agreement with available laboratory experimental data. 
%The simulated spectra of the largest molecule in our set (C$_{55}$H$_{52}$) show qualitative resemblance with the astronomical UIE spectra. 

In order to understand the type of vibrational motions associated with each simulated emission bands, the calculated absorption lines are mapped back to the corresponding simulated emission spectra. 
Through the use of a simple quantitative analysis of cartesian components of displacement vector of all atoms for each normal mode (IR or Raman active), we are able to separate the contributions of aromatic and aliphatic parts. Results show that the major regions of IR spectra affected by aliphatic groups are 3$-$4, 6$-$7, 7$-$8, and 8$-$9 $\mu$m.    Among the last three regions, the region of 7$-$8 $\mu$m contains the largest number of aliphatic vibrations. This is the region of spectra that 8 $\mu$m broad plateau feature is observed in astronomical sources and we suggest that aliphatic side groups play a major role in the formation of the 8 $\mu$m plateau feature. 

The  theoretical scheme described in this work can be very useful in the interpretation of the astronomical spectra and the identification of the structure of the molecular carrier of the UIE bands. In this paper, we have limited ourselves to the study of molecules with an aromatic core and aliphatic branches.  In order to further explore the MAON model, we will study more irregular structures with random aromatic rings connected by randomly oriented aliphatic branches.  Since such structures are more complex, larger computational resources are required.  We hope this work will motivate further theoretical studies of large, amorphous organic compounds for a better understanding of the origin of the astronomical UIE bands.

%The appearance of such broaden feature is explained by contributions of aliphatic groups. %To link the vibrational analysis to astronomical interstellar observations  the IR emission spectra of each species has been simulated at $T=500$ K by applying the Drude thermal excitation broadening model. 

%Without considering the electronic excitation, our combined approach established that the aliphatic groups vibrations and their thermal populations have key roles in merging the PAH bands and forming the broad features in IR emission spectra (especially around the 8 $\mu$m). We also recommend the addition of the aliphatic groups as a computationally economic strategy to enlarge the molecular size and the vibrational modes thereof. The data presented in Table 3 indicates that both C$_{55}$H$_{52}$ (mixed aromatic/aliphatic) and C$_{80}$H$_{24}$ (PAH) have almost the same number of IR active modes in different region of spectrum, however the former introduces 382 electrons and the latter (PAH) 504 electrons to the density functional theory calculations. 

{\flushleft \bf Acknowledgment~}

This work was partially supported by the Research Grants Council of the Hong Kong Special Administrative Region, China (project no. HKU 7031/10P).

\clearpage

\clearpage

\begin{figure}
\epsscale{0.8}
\plotone{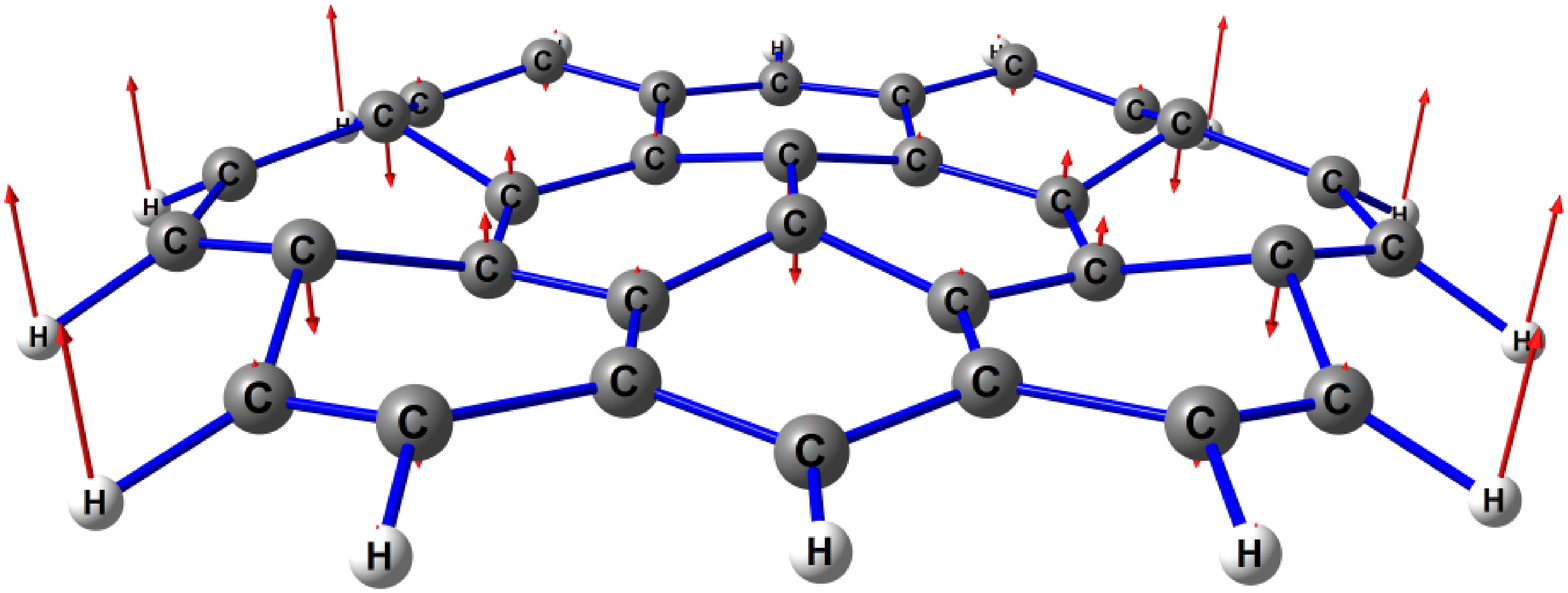}
\caption{A snapshot of animation showing the vibrational motion associated with the 15.59 $\mu$m feature of ovalene (C$_{32}$H$_{14}$).  The direction and magnitude of motions are indicated by arrows.  (A full video is available in the online journal)}
\label{pah_movie}
\end{figure}

\clearpage

\begin{figure}
\plotone{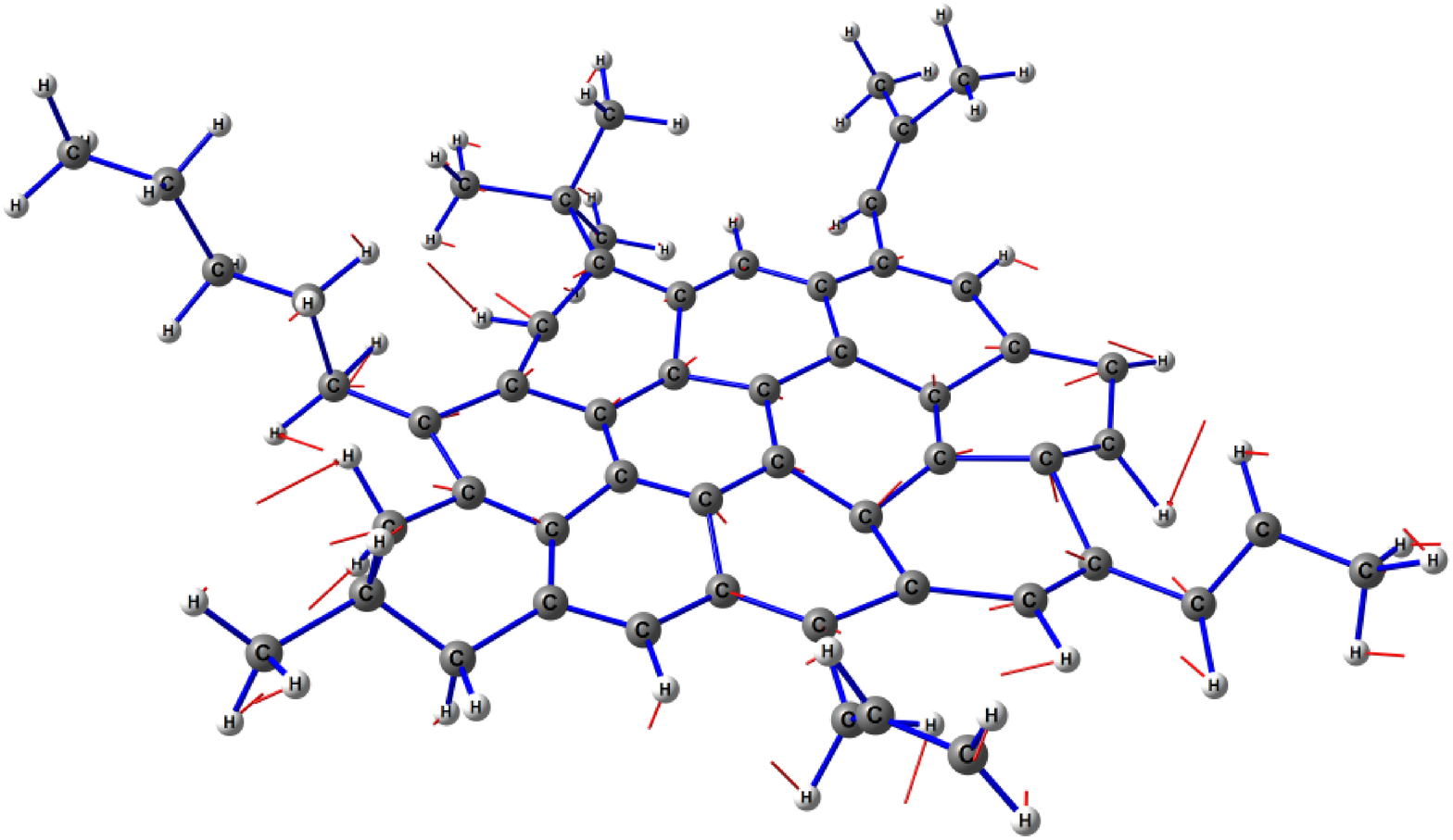}
\caption{A snapshot of  animation showing the vibrational motion associated with the 15.53 $\mu$m feature of C$_{55}$H$_{52}$. The direction and magnitude of motions are indicated by arrows.  (A full video is available in the online journal)}
\label{maon_movie}
\end{figure}

\clearpage

\begin{figure}
\plotone{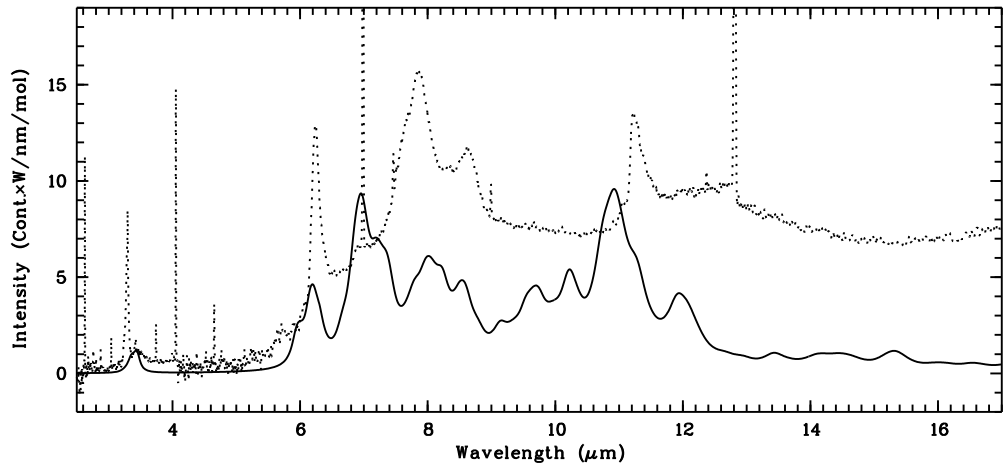}
\caption{Comparison of the simulated emission spectrum of C$_{55}$H$_{52}$ (solid line) with
%(solid line in top panel) with the simulated spectrum (solid line, middle panel) from a collection of nine neutral PAH molecules   (3,4-benzophenanthrene,  chrysene,  phenanthrene, benzanthracene, ovalene, pyrene, coronene, perylene, triphenylene)  \citep[Fig. 5b,][]{cook1998} at 500 K, and the simulated spectrum (solid line, bottom panel) from a collection of 24  neutral, catonic, and anaionic PAH molecules from the PAH database of \citet{boersma2014}.  
the {\it Infrared Space Observatory} spectrum (dotted line) of the planetary nebula BD+30\degr3639 (in units of $F_\lambda$).  The narrow features in the astronomical spectrum are atomic lines.  }
\label{cook}
\end{figure}

\clearpage

\begin{deluxetable}{ccc@{\extracolsep{0.1in}}ccc}
\tablecaption{
The comparison between simulated IR spectrum and
 experimental FT-IR data.  \label{t1} }
\tablewidth{0pt}
\tablehead{
\multicolumn{3}{c}{Experiment$^a$} &
\multicolumn{2}{c}{DFT} &
\colhead{Absolute error$^b$}\\
\cline{1-3}\cline{4-5}
\multicolumn{2}{c}{Position}&
\colhead{Strengths$^c$} &
\colhead{Position}&
\colhead{Strengths$^c$} &
\colhead{($\mu$m)} \\
\cline{1-2}
\colhead{(cm$^{-1}$)} & \colhead{($\mu$m)} &  & \colhead{($\mu$m)}
}
\startdata
626.2&15.96&    0.57&    16.07  &    0.35    &0.11 \\
742.4&13.47&    1.0   &    13.51 &   0.70    &0.04 \\
796.7&12.55&    0.15&    12.59  &   0.15    &0.04 \\
1166.2&8.57&    0.11&    8.76  &   0.17    &0.19 \\
1193.2&8.38&    0.19&    8.44  &   0.24    &0.06  \\
1460.4&6.85&    0.66&    6.90  &   1.0    &0.05 \\
2958.3&3.38&    0.2 &    3.35 &   0.13    &0.03 \\
3028.6&3.30&    0.84&    3.25 &   0.23    &0.05 \\
\enddata
\tablenotetext{a}{The experimental data are the peak positions and
strength of the dominant band at each region of spectra (Sandford et al. 2013).}
\tablenotetext{b}{$|\lambda_{\rm exp}-\lambda_{\rm theo}|$}
\tablenotetext{c}{Normalized to the strongest line.}
\end{deluxetable}

\begin{deluxetable}{cccc}
\tablecaption{Identification of the vibrational modes of emission bands of ovalene.
  \label{ova} }
\tablewidth{0pt}
\tablehead{
\colhead{Bands ($\mu$m)} &
\colhead{Vibrations}
}
\startdata
3.26	&C-H stretching\\
6.18	&C=C stretching coupled with in-plane C-H bending\\
6.59	&C=C stretching coupled with in-plane C-H bending\\
7.19	&C=C stretching coupled with in-plane C-H bending\\
7.80	&C=C stretching coupled with in-plane C-H bending\\
8.22	&C-H in-plane bending\\
8.77	&C-H in-plane bending\\
9.54	&[C=C] in-plane rocking mode unit coupled with  in-plane C-H bending\\
11.11	&C-H out-of-plane bending  coupled with out-of-plane ring deformation\\
11.78	&C-H out-of-plane bending\\
13.04	&C-H out-of-plane bending / In-plane ring deformation\\
14.14	&In-plane ring deformation\\
15.59	&C-H out-of-plane bending  coupled with out-of-plane ring deformation\\
17.94	&C-H out-of-plane bending  coupled with out-of-plane ring deformation\\
\enddata
\end{deluxetable}

\end{document}